\documentstyle[prl,aps,preprint]{revtex}

\begin{document}
\draft

\preprint{\vbox{To be published in {\it Physical Review D} 
\hfill La Plata-TH-95-21}\\}

\title
{Fermion Condensates of massless $QED_2$ at Finite Density
in non-trivial Topological Sectors}
\author{H. R. Christiansen \thanks{CONICET, Argentina.\ \
e-mail address: hugo@venus.fisica.unlp.edu.ar}
{}~and F. A. Schaposnik\thanks{Investigador
CICBA, Argentina. \ \ e-mail address: fidel@athos.fisica.unlp.edu.ar}\\
{\normalsize\it Departamento de F\'\i sica, Universidad
Nacional de La Plata}\\ {\normalsize\it C.C. 67, (1900) La Plata,
Argentina}}

\maketitle

\begin{abstract}
\normalsize
Vacuum expectation values of products of local bilinears
$\bar\psi\psi$   are computed in massless $QED_2$ at finite density.
It is shown that
chiral condensates exhibit an oscillatory inhomogeneous behaviour
depending on the chemical potential.
The use of a path-integral approach clarifies the connection
of this phenomenon with the topological structure
of the theory.
\end{abstract}
\vskip 2cm

\newpage
\pagenumbering{arabic}
\section{Introduction}

Fermion condensates play an important role in
particle physics and cosmology. In particular, they are relevant
in connection with chiral symmetry breaking \cite{NJL},
a phenomenon related to the structure of the QCD vacuum
\cite{condensates}.

Recently, it has been found  in the large $N_C$ limit \cite{Rub},
that the order parameter for chiral symmetry, $<\bar \psi \psi>$,
is at high fermion densities inhomogeneous and anisotropic so that
the ground state of quark matter
has the structure of a standing
wave with respect to the order parameter.

At present, it is well understood that  the
fermion condensate 
$<\bar \psi(x) \psi(x)>$ picks
contributions from non-trivial topological sectors. Then, in order to
decide whether the breakdown of chiral symmetry occurs, one
should consider instanton effects on the fermion
condensate.

Two-dimensional models like $QED_2$ (the Schwinger model) and $QCD_2$
 provide
a natural ground to study these phenomena since, although
simplified, the basic features (chiral
symmetry breaking, non-trivial topological sectors, etc) are
still present and exact calculations can be in many cases performed.
In this context, we calculate in the present work
vacuum expectation values
of products of local bilinears, $\bar \psi \psi$, at finite
density for the Schwinger
model.  We employ a path-integral approach which leads in a very
simple way to exact results and has shown to be very
adequate for studying non-Abelian extensions
\cite{Fid}-\cite{Fidf}. Our conclusions go
in the same direction as those in \cite{Rub} and
extend the results obtained in \cite{Kao}. In particular we are able to
show that multipoint chiral
condensates exhibit an oscillatory inhomogenous
behavior depending on the
chemical potential, as it is the case in the $1/N_c$ results for $QCD_4$
\cite{Rub}. The  lack of translation
invariance  manifests through  a dependence on differences
as well as sums of
spacial coordinates. This last result
corrects the  oscillatory behaviour computed approximately
in \cite{Kao} for the two-point correlator in the Schwinger
model.

Let us start by
observing that Quantum Field Theories at finite fermion density
can be studied
by introducing a chemical
potential leading to a quantum theory
in the presence of a classical background charge
distribution \cite{HW}-\cite{Actor}.
Concerning topological contributions,
Bardakci and Crescimanno \cite{BC} proposed a natural way to take
into account topologically
non-trivial configurations, which is suitable for the study of the
fermion condensates within the path-integral formulation. In this
approach,  one decomposes a given
gauge field belonging to
the $n^{th}$ topological sector in the form

\begin{equation}
A_\mu(x) = { A}_\mu^{(n)}(x) + a_\mu(x)
\label{uno}
\end{equation}
with $ A_\mu^{(n)}$ a fixed configuration carrying the (whole)
topological charge $n$
and $a_\mu$ the path-integral variable, which accounts for
quantum fluctuations and belongs to the trivial topological
sector. In this way, calculations easily workable in
the $n=0$ background
(as the evaluation of Fujikawa jacobians after a fermionic
change of variables using techniques  requiring a
compact manifold) can be handled without problems and, at the same
time, the contribution of topologically non-trivial sectors is properly
taken into account.

\section{The model}
We start from the two dimensional (Euclidean) Lagrangian for $QED_2$

\begin{equation}
{\cal L} = \bar\psi (i\!\!\not\!\partial + \not\!\! A )\psi - \frac{1}{4e^2}
F_{\mu \nu}^2,
\label{lag}
\end{equation}
where our  $\gamma$ matrices are taken as the Pauli matrices,
$\gamma_0 = \sigma_1$, $\gamma_1 = -\sigma_2$
so that $\gamma_\mu \gamma_5 =   i\epsilon_{\mu \nu} \gamma_\nu$.

The partition function ${\cal Z}$ for the model is
\begin{equation}
{\cal Z} =  \int D\bar \psi D \psi DA_\mu \exp (- \int d^2x {\cal L})
\label{par}
\end{equation}

In order to include fermion density effects, it will be convenient
to add still another vector field $A_\mu^b$ which will
allow for the introduction of
the chemical potential in a simple way. This field describes
an {\em external}
charge density acting
on the quantum system.
Whenever  $A_\mu^b$ is
taken as $i$ times the chemical potential \cite{Actor}, it will represent
a uniform charge background. If one  first considers a finite length
($2 L$)
distribution and then takes the $L\rightarrow\infty$ limit,
translation symmetry breaking becomes apparent and at the same time
ambiguities in the definition of the finite density theory are avoided
(see ref.\cite{fks}).

Now we can write the gauge field
in Lagrangian (\ref{lag}) in the form
\begin{equation}
A_\mu (x) = { A}_\mu^{(n)}(x)  + A_\mu^b(x) + a_\mu(x)
\label{dos}
\end{equation}
 Hence, the Lagrangian in the $n^{th}$ topological sector reads

\begin{eqnarray}
{\cal L}^{(n)}  & = &
\bar\psi (i\!\!\not\!\partial + \not\! a + {\not\!\! A}^{(n)} +
{\not\!\! A}^b) \psi
- \frac{1}{4e^2}( f_{\mu \nu}^2 + {{ F}_{\mu \nu}^{(n) 2}}
+ 2 f_{\mu \nu} {{F}_{\mu \nu}^{(n)}}
+ \nonumber \\
& &  {(F_{\mu \nu }^b)}^2 + 2 f_{\mu \nu}F_{\mu \nu}^b +
2 F_{\mu \nu}^{(n)}F_{\mu \nu}^b) + {\cal L}_c
\label{lagu}
\end{eqnarray}
In eq.(\ref{lagu}) we have added a counterterm ${\cal L}_c$ which will
be conveniently chosen to cancel out a divergency arising when the
 background
$A_\mu^b$ is taken as a constant to introduce the chemical
potential (see below).

The partition function now reads
\begin{equation}
{\cal Z} =
\sum_n \int D\bar \psi D \psi Da_\mu \exp (- \int d^2x {\cal L}^{(n)})
\label{par1}
\end{equation}
where we have written the path-integral as a sum over all
topological sectors.

As it is well-known, in two dimensions
fermions can be completely decoupled from
gauge fields by an appropriate chiral and gauge transformation of
the fermion fields. At
the quantum level, the corresponding change in the fermionic variables
is accompanied by a non-trivial Fujikawa jacobian \cite{Fujikawa}.
As it can be inferred from the connection between the chiral anomaly (and
 the index theorem for the Dirac operator) with the change in the
fermionic path-integral measure,
calculation of the jacobian can be more easily performed
in compact space-time manifolds.
The latter implies that only decouplings from topologically trivial gauge
fields should be considered. With this in mind we perform the following
change in the fermionic variables

\begin{eqnarray}
\psi & = & \exp (\gamma_5 \phi + i \eta + \gamma_5 \phi^b + i \eta^b) ~
\chi
\nonumber
\\
\bar \psi & = & \bar \chi
{}~ \exp (\gamma_5 \phi - i \eta + \gamma_5 \phi^b - i \eta^b)
\label{trafo}
\end{eqnarray}
 according to a standard decomposition of the vector fields
\begin{equation}
a_\mu = - \epsilon_{\mu \nu} \partial_\nu \phi + \partial_\mu \eta
\label{a}
\end{equation}
and

\begin{equation}
A_\mu^b = - \epsilon_{\mu \nu} \partial_\nu \phi^b + \partial_\mu \eta^b
\label{a1}
\end{equation}

\noindent With this change of variables the only field that
remains coupled to the fermions is $A_\mu^{(n)}$, a
classical configuration
carrying all the topological charge.
In terms of the new variables,  Lagrangian (\ref{lagu}) becomes
\begin{equation}
{\cal L}^{(n)}  =
\bar\chi (i\!\!\not\!\partial  + {\not\!\! A}^{(n)})\chi
-\frac{1}{4e^2} {\left( F_{\mu \nu}^{(n)} \right)}^2
- \frac{1}{2e^2}\left(
(
\Box
 \phi)^2 + \epsilon_{\mu\nu}
{F}_{\mu\nu}^{(n)}
\Box
\phi \right ) + {\cal L}_c
\label{ult}
\end{equation}
where we have already taken $A_\mu^b$ as a constant, in order to
introduce the chemical potential.

Eq.(\ref{ult}) describes the effect of the
change of variables merely at the classical
level. At the quantum level, within the path-integral approach,
 one has to take
into account the change in the fermionic measure, i.e., the Fujikawa
jacobian associated with transformations (\ref{trafo}).
The evaluation of this jacobian is standard and we just quote the result
\cite{Fid}
\begin{equation}
J  =  \exp \left(\frac{1}{2\pi}\int d^2x ~\phi
\Box (\phi + 2 \phi^{(n)})
\right) \times
  \exp \left(\frac{1}{2\pi}\int d^2x ~\phi_b \Box (\phi + 2 \phi^{(n)})
\right)
\label{jac}
\end{equation}
where $\phi^{(n)}$ is  accordingly defined by
\begin{equation}
A_\mu^{(n)} = - \epsilon_{\mu \nu} \partial_\nu \phi^{(n)}
\label{top}
\end{equation}

\noindent Concerning the Jacobian associated with the change  of the
bosonic variables
(eqs.(\ref{dos}),(\ref{a})),
\begin{equation}
Da_\mu = J_{bos} D\phi D\eta ~,
\label{cam}
\end{equation}
it can be identified with  the Faddeev-Popov determinant
for the $\eta = 0$ (Lorentz) gauge, $ J_{bos} ={ \rm det}\Box$,
and it will be
ignored in what follows.

To relate $A_\mu^b$ with
 the chemical potential $\mu$, note that fixing the fermion
number density through a term
\begin{equation}
{\cal L}_{chem} = -i \mu \bar \psi \gamma_0 \psi
\label{chem}
\end{equation}
corresponds to the choice
\begin{equation}
A_\nu^b = -i \mu \delta_{\nu 0}
\label{cons}
\end{equation}
or, equivalently,
\begin{equation}
\phi_b = i \mu x_1\ \ \ \ \eta_b = 0
\label{ficon}
\end{equation}
Notice that, as has been thoroughly analysed by Actor \cite{Actor},
$A_\mu^b$ does not correspond to a pure gauge. Were it not so,
the introduction  of a chemical potential would not have physical
consequences. For the same reason, one cannot gauge away ${\cal L}_{chem}$
from the lagrangian by means of the alternative choice: $\eta_b=-i\mu x_0$,
$\phi_b=0$. In fact, this transformation would correspond to an unbounded (in the temporal axis) gauge one. Although transformation (\ref{ficon})
is unbounded as well, as mentioned above one can handle this
problem as it is usually done: puting the system in a spatial
box, introducing counterterms and taking the infinite limit for 
its length at the end of the calculations.
In this way the usual divergency associated to the electromagnetic energy carried by fermions can be eliminated by an appropiate counterterm \cite{fks}.
In our approach the divergency manifests through the term
 $ \phi_b \Box \phi^{(n)}$ in eq.(\ref{jac}).
As stated above, this infinite contribution is cancelled out by an
appropriate choice of ${\cal L}_c$.
This counterterm is the Lagrangian counterpart of the
usually employed in the Hamiltonian approach to handle this problem
\cite{Kao,fks}.
In the canonical QFT
this is tantamount to
a redefinition of creation and annihilation operators which is
equivalent to a shift in the scale used to measure excitations.

Putting all this together, the partition function of the model,
can be written in the form
\begin{equation}
{\cal Z} = {\cal N}\sum_n
\int D \bar \chi D \chi D\phi ~ \exp (- S_{eff}^{(n)})
\label{ef}
\end{equation}
where $ S_{eff}^{(n)}$ is the effective action in each
topological sector,

\begin{eqnarray}
S_{eff}^{(n)} & = &\int d^2x ~ \bar\chi (i\!\!\not\!\partial  +
{\not\!\! A}^{(n)})\chi -
\frac{1}{2e^2}\int d^2x \left(
(\Box
 \phi)^2 + \epsilon_{\mu\nu}  { F}_{\mu\nu}^{(n)} \Box
\phi \right ) +  \nonumber \\*[2 mm]
& & -\frac{1}{4e^2} \int d^2x {(F_{\mu \nu}^{(n)})}^2
-\frac{1}{2\pi}\int d^2x ~ \phi \Box (\phi + 2 \phi^{(n)})
 \label{lar}
\end{eqnarray}

Note that with the choice of ${\cal L}_c$ discussed above,
the effective action written in terms of the "decoupled" fermions
does not depend  on the chemical potential $\mu$. Nevertheless,
$\mu$ reappears when  computing correlation functions of fermion
fields, once $\bar \psi$ and $\psi$ are written in terms of
the decoupled fields $\bar \chi$ and $\chi$ through eq.(\ref{trafo}).

Of course, the fermionic integral in eq.(\ref{ef}) is
the determinant of the Dirac operator in the background
of a gauge field carrying topological charge $n$. Now, as it is well-known
\cite{BC}, the Dirac operator has,
for $n>0$ ($n<0$), $n$ positive (negative) chirality zero modes so that
actually none but the
 $n=0$ sector does contribute to ${\cal Z}$. However, in computing
 v.e.v.'s of products of fermion bilinears,
Grassman coefficients accompanying  zero-modes render non-trivial
certain
path-integrals
 in a given topological sector. This happens according to the number of
bilinears appearing in the v.e.v.,  and then
only this sector will contribute (see next section).

\section{Correlation functions of fermion bilinears}

We are now ready to study the behavior of chiral condensates and
their dependence on the chemical potential $\mu$. To
this end, let us define  the chiral charge changing  correlators
\begin{equation}
s_+(w) = \bar \psi_+ \psi_+ (w)
\label{smas}
\end{equation}
\begin{equation}
s_-(w) = \bar \psi_- \psi_- (w)
\label{smen}
\end{equation}

\noindent where $\psi_+$ and $\psi_-$ are the  right-handed and left-handed
components of the Dirac spinors.
Thus, the fermion condensate $<\bar \psi \psi(w)>$ is the sum of the
v.e.v.'s
of the composites
defined in eqs.(\ref{smas})-(\ref{smen}).

Notice that this correlator could be naively expected to vanish as a direct
consequence of the chiral invariance of massless $QED_2$.
However, as explained above, when one is to compute
$<s_{+}(w)>$ ($<s_{-}(w)>$)
using the partition
function given by eqs.(\ref{ef})-(\ref{lar}), a non zero value
is found. In fact, only the $n=1$ ($n=-1$)
sector will contribute to the sum over topologically
non trivial sectors. One then has

\begin{eqnarray}
<s_\pm(w)> & = & \frac{1}{{\cal Z}_0}\int
 D \bar \chi D \chi D\phi ~ \bar \chi_\pm(w) \chi_\pm(w)
\exp \left( \pm 2(\phi(w) + \phi_b(w)) \right)
\nonumber \\
& & \exp (- S_{eff}^{(\pm 1)})
\label{hu}
\end{eqnarray}
where ${{\cal Z}_0}$, the only non-vanishing contribution
to ${\cal Z}$, arises
from the $n=0$ sector.
After some algebra, eq.(\ref{hu}) becomes

\begin{eqnarray}
<s_\pm(w)> & = &\frac{1}{{\cal Z}_{F}}\int
 D \bar \chi D \chi ~ \bar \chi_\pm(w) \chi_\pm(w)
\exp(-\int d^2x  ~ \bar\chi (i\!\!\not\!\partial  +
{\not\!\! A}^{(\pm 1)})\chi)
\nonumber \\
& & \frac{1}{{\cal Z}_B}\int D\phi \exp \left( \pm 2(\phi(w) + i \mu w_1)
\right)
\times \exp(-\frac{1}{4e^2} ({F_{\mu \nu}^{(\pm 1)}})^2)
\nonumber \\
& & \exp (\frac{1}{2\pi} \int d^2x (\phi \Box \phi + 2
 \phi \epsilon_{\mu \nu}\partial_\mu A_\nu^{(\pm 1)}) \label{quin}
\end{eqnarray}
with
\begin{equation}
Z_F = {\rm det} i\!\!\not\!\partial
\label{det}
\end{equation}
and
\begin{equation}
Z_B = {\rm det}^{-1/2}\left(\Box(\Box - \frac{e^2}{4\pi}) \right)
\label{detb}
\end{equation}

It can be easily proved in the $\mu = 0$ case that $<s_+>_{+1}^{\mu = 0}$
and $<s_->_{-1}^{\mu = 0}$ coincide. This in turn implies that
\begin{equation}
<\bar \psi \psi(w)>^{\mu = 0} = 2 <s_+>_{+1}^{\mu = 0}
\label{turn}
\end{equation}
Now, from eq.(\ref{quin}) we can see that
\begin{equation}
<s_\pm>_{\pm 1} = \exp(\pm 2i \mu w_1) <s_\pm>_{\pm 1}^{\mu = 0} ~,
\label{uf}
\end{equation}
hence, one finally gets

\begin{equation}
<\bar \psi \psi(w)>^{\mu \ne 0} =
\cos (2\mu w_1) <\bar \psi \psi(w)>^{\mu = 0}
\label{ufi}
\end{equation}
In order to have an explicit formula
for eq.(\ref{ufi}), let us recall that
\begin{equation}
 <s_+>_{+1}^{\mu = 0} =  - \frac{m}{4\pi} e^{\gamma}
\label{revi}
\end{equation}
where $\gamma$ is the Euler constant and $m = e^2/\pi$ is the mass
of the effective boson\cite{CKS}.
Eq.(\ref{ufi}) coincide with that presented in \cite{Kao}
using operator bosonization rules. Appart from the simplicity
of our derivation in the path-integral framework, it should be
stressed that it is within this approach that the role of different
topological sectors becomes apparent. This can be put in evidence
in the calculation of $N$ point correlators which will be discussed
in the remaining of this section.

According to the previous discussion, we consider the
contribution of the $n^{th}$ topological sector to the $N$
point correlation function

\begin{eqnarray}
<\prod_{i=1}^{N} s_\pm(w_i)>_{\pm n} ^{\mu \ne 0}& = &
\frac{1}{{\cal Z}_0}\int D\bar \chi D\chi D\phi
\prod_{i = 1}^N \bar \chi P_\pm \chi(w_i) \times \nonumber \\
& & \exp( \pm 2 \sum_{i = 1}^N \theta(w_i) - S_{eff}^{(\pm n)})
\label{pri}
\end{eqnarray}
Here $<~>_{\pm n}$ means that the v.e.v. is computed in the
$\pm n^{th}$-topological sector;  $P_+$ ($P_-$) is the projector on
the right-handed (left-handed) subspace
\begin{equation}
P_\pm = \frac{1}{2} ( 1 \pm  \gamma_5)
\label{pro}
\end{equation}
and we have defined
\begin{equation}
\theta (w) = \phi(w) + \phi^b(w)
\label{pru}
\end{equation}
Now,  due to Grassman integration rules, it can
be easily proved that the only non vanishing contribution to these
correlators may arise when the number of insertions equals the absolute
value of the topological charge, ie: $|n|= N$ \cite{Naon}. Performing the
decoupling of fermions as
before we finally get the following expression

\begin{equation}
<\prod_{i=1}^{N} s_\pm(w_i)>_{\pm N}^{\mu \ne 0} =
F_{\pm N}(w_1,w_2, ...,w_N) B_{\pm N}(w_1,w_2, ...,w_N)
\label{pra}
\end{equation}
Here
\begin{equation}
F_{\pm N} = Z_F^{-1} \int D\bar\chi D\chi\
\prod_{i = 1}^N \bar \chi_\pm \chi_\pm(w_i)
\exp(-\int d^2x \bar \chi  (i\!\!\not\!\partial  +
{\not\!\! A}^{(\pm N)})\chi )
\label{pas}
\end{equation}
and
\begin{eqnarray}
 B_{\pm N} & = & Z_B^{-1}\int D\phi\ \exp(\pm 2 \sum_{i = 1}^N
\theta(w_i) )
\exp (\frac{1}{2 e^2} \int d^2x \phi \Box (\phi +
2\phi^{(\pm N)} )) \times \nonumber \\
& &  \exp \left( \frac{1}{2 e^2} \int d^2x ( (\Box \phi)^2 +
2 \Box \phi^{(\pm N)} \Box \phi + {(\Box \phi^{(\pm N)})}^2 \right)
\label{cu}
\end{eqnarray}
We thus see that the dependence of the correlator on the chemical
potential factorizes, resulting in

\begin{equation}
<\prod_{i=1}^{N} s_\pm(w_i)>_{\pm N}^{\mu \ne 0} =
\exp(\pm i\mu \sum_{i = 1}^N w_{i 1}) \times
<\prod_{i=1}^{N} s_\pm(w_i)>_{\pm N}^{\mu = 0}
\label{pp}
\end{equation}
where

\begin{equation}
<\prod_{i=1}^{N} s_\pm(w_i)>_{\pm N}^{\mu = 0} = (-m e^\gamma/4\pi)^N
\exp (-2 \sum_{i > j} K_0(m \vert w_i - w_j \vert) )
\label{min}
\end{equation}

\noindent are the general minimal correlation functions for $\mu = 0$
\cite{Naon,Steele}.

We are now ready to go further and study the so-called
{\it non-minimal} \cite{BC},\cite{Steele,mnt93} correlation
functions which will be needed in order to compute multipoint vacuum
condensates.
We start considering  the complete two point composite

\begin{eqnarray}
<\bar \psi \psi(x) \bar \psi \psi(y)> & = &
<s_+(x)s_+(y)> + <s_-(x)s_+(y)> \nonumber\\
& + &  <s_+(x)s_-(y)>
+ <s_-(x)s_-(y)>
\label{sumi}
\end{eqnarray}
In the $n= 0$ topological sector the unique contribution to eq.(\ref{sumi})
comes precisely from the simplest { non-minimal}
v.e.v. which after fermion decoupling becomes

\begin{eqnarray}
& & <s_-(x)s_+(y)>^{\mu \ne 0} +  <s_+(x)s_-(y)>^{\mu \ne 0}
= \nonumber \\
& & \big( \exp(-2i\mu (x_1 - y_1)) \times
 <\bar \chi_+ \chi_+(x) \bar \chi_- \chi_-(y)>_F + \nonumber \\
& & \exp(2i\mu (x_1- y_1)) <\bar \chi_-\chi_-(x) \bar \chi_+
\chi_+(y)>_F \big)
\int D\phi\ e^{\int d^2x(\frac{1}{2e^2}(\Box\phi)^2+\frac{1}{2\pi}
(\phi\Box\phi))}
\label{qui}
\end{eqnarray}

This gives a contribution to $<\bar \psi \psi(x) \bar \psi \psi(y)>$
in the $n=0$ topological sector, of the form
\begin{equation}
<\bar \psi \psi(x) \bar \psi \psi(y)>^{\mu \ne 0}_{n=0} =
\cos (2\mu (x_1 - y_1)) <\bar \psi \psi(x) \bar \psi
\psi(y)>^{\mu = 0}_{n=0}
\label{qq}
\end{equation}
There are two remaining contributions arising from the $n = \pm 2$
sectors.
They follow directly from eq.(\ref{pp})

\begin{eqnarray}
<s_+(x)s_+(y)>^{\mu \ne 0}_{+2} & = & \exp(-2i\mu (x_1 + y_1))
<s_+(x)s_+(y)>^
{\mu = 0}_{+2} \nonumber \\
<s_-(x)s_-(y)>^{\mu \ne 0}_{-2} & = & \exp(2i\mu (x_1 + y_1))
<s_-(x)s_-(y)>^
{\mu = 0}_{-2}
\label{ff}
\end{eqnarray}
Again
\begin{equation}
<s_+(x)s_+(y)>^
{\mu = 0}_{+ 2} = <s_-(x)s_-(y)>^
{\mu = 0}_{-2} = \frac{1}{2}<\bar \psi \psi(x) \bar \psi
\psi(y)>_{\vert n \vert = 2}^{\mu \ne 0}
\end{equation}
so that we finally have
\begin{eqnarray}
& & <\bar \psi \psi(x) \bar \psi \psi(y)>^{\mu \ne 0} = \cos (2\mu (x_1
- y_1)) \times
<\bar \psi \psi(x) \bar \psi \psi(y)>^{\mu = 0}_{n = 0} + \nonumber \\
 & & \cos (2\mu (x_1
+ y_1)) \times
<\bar \psi \psi(x) \bar \psi \psi(y)>^{\mu = 0}_{\vert n \vert = 2}
\label{yy}
\end{eqnarray}
At this point some remarks are in order:
\vspace{0.2 cm}

\noindent (i) Once again the  topological structure of the theory
manifests, making possible
the discrimination of the contributions
to  fermion bilinear correlation functions from each topological
sector.\vspace{ 0.1 cm}

\noindent (ii) Eq.(\ref{yy})
exhausts {\it all}  topological contributions.
\vspace{ 0.1 cm}

\noindent (iii) The lack of translation invariance
(which is broken by the
background charge distribution) becomes apparent particularly through the
last term in the r.h.s. of eq.(\ref{yy}), which depends on
the combination $x_1 + y_1$
\vspace{ 0.1 cm}

\noindent (iv) No clustering ansatz has been needed in order to
obtain these results.
\vspace{ 0.1 cm}

Let us finally note that if we put $ y_1 = 0$ in eq.(\ref{yy}) we
get a compact formula analogous to eq.(\ref{ufi}),

\begin{equation}
<\bar \psi \psi(x) \bar \psi \psi(y)>^{\mu \ne 0} = \cos (2\mu x_1)
<\bar \psi \psi(x) \bar \psi \psi(y)>^{\mu = 0}
\label{an}
\end{equation}
This last result can be seen to coincide
with that obtained in ref.\cite{Kao} using  cluster decomposition.
 In fact, the result reported in \cite{Kao} corresponds
just to the trivial topological sector and does not reproduce the
contribution of $n \ne 0$ sectors whenever $x_1 \ne y_1$.

Correlators of a larger number of bilinears can be very simply
obtained following
the same procedure as above. As an example one gets for the $3$-point
correlator,

\begin{eqnarray}
& & \sum_n  <\bar \psi \psi(x) \bar \psi \psi(y) \bar \psi \psi(z)>
_n ^{\mu \ne 0} =
\nonumber \\
& & 2 \cos(2\mu(x_1 + y_1 + z_1))<s_+(x)s_+(y)s_+(z)>^{\mu = 0}_{n=3} +
\nonumber \\
& & 2 \cos(2\mu(x_1 + y_1 - z_1))<s_+(x)s_+(y)s_-(z)>^{\mu = 0}_{n=1} +
\nonumber \\
& & 2 \cos(2\mu(x_1 - y_1 + z_1))<s_+(x)s_-(y)s_+(z)>^{\mu = 0}_{n=1} +
\nonumber \\
& &
2 \cos(2\mu(- x_1 + y_1 + z_1))<s_-(x)s_+(y)s_+(z)>^{\mu = 0}_{n=1}
\label{tres}
\end{eqnarray}
(We have emphasized that the l.h.s. of eq.(\ref{tres}) exhausts
all the topological contributions by explicitely showing the sum over
$n$).

{}From the examples above one can easily infer the structure for the
general $N$ point correlator
\begin{eqnarray}
& & <\bar \psi \psi(w^1) \bar \psi \psi(w^2)
\ldots \bar \psi \psi(w^N)>^{\mu \ne 0} =
\nonumber
\\*[2 mm]
& &
\sum_k
\cos(2\mu(\sum_{i} w^i_1))<s_+(w^1)s_+(w^2)\ldots
s_+(w^N)>_{n = N}^{\mu = 0} +
\nonumber \\*[2 mm]
& &
\sum_k
\cos(2\mu(\sum_{i \ne k} w^i_1 - w^k_1))<s_+(w^1)s_+(w^2)\ldots
s_+(w^{k - 1})
s_-(w^k)
\nonumber \\*[2 mm]
& & s_+(w^{k+1})  \ldots s_+(w^N)>_{n = N-1}^{\mu = 0} +
 \sum_{k,j}
\cos(2\mu(\sum_{i \ne k,j} w^i_1 - w^k_1 - w^j_1))<s_+(w^1)
\nonumber \\*[2 mm]
& &
s_+(w^2)\ldots s_+(w^{k - 1})
s_-(w^k) s_+(w^{k+1}) \ldots
s_+(w^{j - 1})
s_-(w^j)\nonumber \\*[2 mm]
& &  s_+(w^{j+1}) \ldots s_+(w^N)>_{n = N-2}^{\mu = 0}
+ \ldots
\label{larga}
\end{eqnarray}

A compact expression for the non-minimal correlation functions
appearing in the last equation is

\begin{equation}
<\prod_{i=1}^{r} s_{+}(w_i) \prod_{j=1}^{s} s_{-}(w_j)>_{r-s}^{\mu = 0} =
(-m e^\gamma/4\pi)^N
\exp (-2 \sum_{i > j} e_i e_j K_0(m \vert w_i - w_j \vert) )
\label{nomin}
\end{equation}

where $r+s=N$ (for the details see \cite{Steele}).

\vskip 0.5cm

In summary, we have presented the
correlation functions of fermion bilinears
in $QED_2$ at finite density,
using a path-integral approach particularly adequate
for identifying contributions arising from different topological
sectors. We have been able to exactly compute  correlation functions
for an arbitrary number of bilinears, showing its dependence
with the chemical potential.
One of our motivations was a recent work by Deryagin,
Grigoriev and Rubakov  \cite{Rub} where it has been
shown that in the large $N_C$ limit, condensates in $QCD$
are inhomogeneous and anisotropic at high fermion density.

The Schwinger model is a favorite laboratory to test
phenomena which are expected to happen in $QCD_4$. In fact,
an oscillatory inhomogeneous behavior in
$<\bar \psi \psi>$ was discussed in the Schwinger model
\cite{Kao} using operator bosonization. We think
that the path-integral approach employed in the present paper
is more appropriate to make apparent the crucial role that topological
sectors play  in the behavior of condensates
(they are actually responsible for the
non-vanishing of $<\bar \psi \psi>$).
In fact, our analysis implies that
the phenomenon is not just a byproduct of $2$-dimensional peculiarities.

It is striking that the
oscillatory behavior that we
have found, exactly coincides (appart from the anisotropy that of course
cannot be tested in one spacial dimension) with that
described in \cite{Rub} for $QCD_4$.
The structure of the $N$ point correlation functions, given by
eq.(\ref{larga}), shows a non trivial
dependence on spatial coordinates. This makes apparent that
the ground state has, at finite density, an
involved structure which is a superposition of standing
waves with respect to the order parameter.

Several interesting issues  are open for further investigation
using our approach.
One can in particular study in a very simple way the
behavior of condensates at finite temperature. The chiral anomaly
is independent of temperature and plays a central role
in the behavior of condensates
through its connection with the index theorem.  Therefore,
one should expect
(as discussed in ref. \cite{Kao} for $<\bar \psi \psi>$)
that a formula like (\ref{larga}) is valid also for $T > 0$. Of
course, v.e.v.'s at $\mu = 0$ in the r.h.s.
of this equation, should be replaced
by those computed at finite temperature and hence  the issue of zero-
modes in a toroidal manifold should be
carefully examined (see e.g. \cite{Steele}).

Another extension which can be undertaken is related to the study of
massless $QCD_2$. Indeed, the decoupling change of variables at the root
of our approach, can be easily extended for non-abelian gauge
groups and has lead to deep insights in the properties of the model
\cite{Fid1}-\cite{Fidf}.
Finally, it should be worthwhile to consider massive fermions and
compute fermion correlation functions, via a perturbation
expansion in the fermion mass following the approach of
\cite{NNaon}. We hope to report on these problems in a future work.

\vfill

Acknowledgements: One of us (H.R. C.) would like to thank
 C.M. Naon for suggesting
the study of this problem and M.V. Man\'\i as and M. Trobo for
helpful comments.
This work was
supported in part by CICBA and CONICET.


\end{document}